\documentclass[showpacs,aps,preprintnumbers]{revtex4}
\usepackage{graphicx}% Include figure files
\usepackage{dcolumn}% Align table columns on decimal point

\begin{document}

\title{Improved Gell-Mann$-$Okubo Relations and SU(3) Rotation Excitations of Baryon States\footnote{The project supported by National
Physics Project, National Natural Science Foundation of China and
Science Fund of the Chinese Academy of Sciences. }}
\author{Mu-Lin Yan}
\email{mlyan@ustc.edu.cn} \affiliation{ Fundamental Physics
Center, University of Science and Technology of China, Hefei,
Anhui 230026, P.R. China}
\author{Xin-He MENG}
\affiliation{  CCAST(World Laboratory), P.O. Box 8730, Beijing
100080 \\
and \\
Physics Department, Nankai University, Tianjin 300071,
P.R.China\footnote{Mailing address.}}

%\begin{center}
 Published in Commun. Theor. Phys.(Beijing) {\bf 24}
(1995) pp. 435-438; (Received August 3, 1994)

--------------------------------------------------------------------------------------------------------
------------------------------
%\end{center}
%\preprint{USTC-ICTS-02036}

\begin{abstract}
The corrections to the Gell-Mann--Okubo relations of baryon masses
are presented in the SU(3) Skyrme model. These corrections are
calculated up to the second order in flavor breaking at the
skyrmion quantum mechanics. The results are compatible with the
experimental data. They could be regarded as a signal of existence
of the SU(3) rotation excitation states of baryons: 27-let (with
spin 1/2 or 3/2), 10*-let (with spin 1/2) and 35-let (with spin
3/2).
\end{abstract}

\pacs{11.30.Rd, 12.39.Dc, 12.39.Mk, 13.75.Gx}

\maketitle

\section*{}

The Gell-Mann--Okubo relations (GOR) of baryon
masses\cite{Gell-Mann} were originally formulated in terms of a
perturbative treatment of flavor breaking in SU(3) group theory.
In the history, the success of GOR (or SU(3) theory) led to the
quark conception and deep understanding for the strong
interactions. In recent years a new (GOR type) baryon mass formula
including octet and decuplet masses has been deduced\cite{Mor}
using only general properties of QCD. Also some effective theories
of QCD derived\cite{Jen} new mass formulas improving GOR.
 According
to GOR, the baryon-octet and baryon-decuplet mass relations can be
written as
\begin{eqnarray}\label{2.1}
2(M_N+M_\Xi )&=&3M_\Lambda +M_\Sigma+\delta m_8, \\
M_\Delta-M_{\Sigma^*}&=&M_{\Sigma^*}-M_{\Xi^*}+\delta m_{10}^{(1)}
=M_{\Xi^*}-M_\Omega+\delta m_{10}^{(2)}.
\end{eqnarray}
Here as $\delta m_8=\delta m_{10}^{(1)}=\delta m_{10}^{(2)}=0$,
equations (1) and (2) are the standard GOR. However, in the real
world
\begin{equation}\label{2.4}
\delta m_8=-26 MeV,\;\;\;\delta m_{10}^{(1)}=4.5MeV,\;\;\;\delta
m_{10}^{(2)}=9.6MeV,
\end{equation}
which stand for the deviations of GOR from experimental data. The
motivation of our studies in the present paper is to calculate
$\delta m_8,\;\;\delta m_{10}^{(1)}$ and $\delta m_{10}^{(2)}$ in
terms of the Skyrme model\cite{Sky,Gua} analytically.

In the Skyrme model the baryon octet and baryon decuplet emerge as
topological solitons (i.e., skyrmions) in the SU(3)$\times$SU(3)
current algebraic chiral Lagrangians. It is believed that this
chiral soliton model provides a reasonable dynamical mechanism for
the mass splitting of SU(3) baryons. In actual factor, through
calculations of the first order perturbation to the masses of
baryons in the SU(3) Skyrme model, one can re-establish
GOR\cite{Gua}. Thus it could be expected that the high order
correction calculations in the perturbations of the skyrmion
quantum mechanics will show the the values of $\delta
m_8,\;\;\delta m_{10}^{(1)}$ and $\delta m_{10}^{(2)}$. In the
present paper we will complete the calculations of the second
order corrections which should be the leading order to $\delta
m_8,\;\;\delta m_{10}^{(1)}$ and $\delta m_{10}^{(2)}$, and learn
some new physical implications from them.

The Hamiltonian of the SU(3) skyrmion quantum mechanics in the
SU(3) collective coordinate space is\cite{Kan} (we use the
notations of ref.\cite{Kan} hereafter)
\begin{eqnarray}
H&=&H_0+H', \\
H_0&=&M_s+{1 \over 2b^2}(\sum\limits_{i=1}^{8}L_iL_i-R_8^2)+
{1\over 2}({1\over a^2}-{1\over b^2})\sum\limits_{A=1}^{3} R_A R_A
+{2\delta \over \sqrt{3}}F_\pi R_8,\\
H'&=&m[1-D_{88}^{(ad)}(A)],
\end{eqnarray}
where $M_s$ is a classical soliton's mass, $a^2$ and $b^2$ are the
soliton's moments of inertia, $m,\;\;F_\pi$ and $\delta$ are
constants and parameters in the model, $D_{88}^{(ad)}(A)$ denotes
the regular adjoint representation functions of SU(3), and $[L_i,
L_j]=if_{ijk}L_k,\;\;[R_i, R_j]=-if_{ijk}R_k,\;\; [L_i, R_j]=0$.
In Eq.(4) $H_0$ serves as the unperturbed Hamiltonian and $H'$ as
the perturbative part. It is easy to see that $H_0$ is diagonal
and the eigen-wave functions for $H_0$ are\cite{Kan}
\begin{equation}
|^\lambda_{\mu\nu}\rangle=(-1)^{s+s_z}\sqrt{\lambda}D_{\mu\nu}^{(\lambda)}
(A),\hskip0.1in\;\;\;\mu=(^{IY}_{I_3}),\hskip0.1in
\;\;\;\;\nu=(^{s\;1}_{-s_z}).
\end{equation}
$|^8_{\mu\nu}\rangle$ and $|^{10}_{\mu\nu}\rangle$ correspond to
baryon octet and baryon decuplet respectively. As one computes the
matrix elements of $\langle H' \rangle $ a useful formula is as
follows:
\begin{equation}
\langle^{\lambda_2}_{\mu_2\nu_2}| D_{\mu\nu}^{(\lambda)}
(A)|^{\lambda_1}_{\mu_1\nu_1}\rangle=(-1)^{s_1+s_{1z}+s_2+s_{2z}}\sqrt{{\lambda_1
\over \lambda_2} }\sum\limits_{\gamma}\left (
   \begin{array}{lcr}
   \lambda_1 & \lambda & \lambda_{2\gamma}\\
   \mu_1     & \mu     & \mu_2
   \end{array} \right )
\left (
   \begin{array}{lcr}
   \lambda_1 & \lambda & \lambda_{2\gamma}\\
   \nu_1     & \nu     & \nu_2
   \end{array} \right )
\end{equation}
with a standard notation for the Clebsch-Gordan (CG) coefficients.
Here $\gamma$ distinguishes the independent irreducible
representations occurring in the reduction $(\lambda) \otimes
(\lambda_1)\rightarrow (\lambda_2)$. The mass of the baryon for
$|k\rangle \equiv |^\lambda_{\mu\nu}\rangle $ can be calculated in
perturbation
\begin{equation}
M_k=E_k^{(0)}+ E_k^{(1)}+E_k^{(2)}+\cdots ,
\end{equation}
where
\begin{eqnarray}
E_k^{(0)}&=&\langle k|H_0|k \rangle, \\
E_k^{(1)}&=&\langle k|H'|k \rangle,\\
E_k^{(2)}&=&\sum\limits_{n\neq k}{|\langle n|H'|k \rangle |^2
\over E_k^{(0)}-E_n^{(0)}}
\end{eqnarray}
$E_k^{(0)}$ and $E_k^{(1)}$ have been known in
literatures\cite{Kan}. Our object is to calculate $E_k^{(2)}$
which is related to non-diagonal matrix elements of $D^{(8)}_{8,8}
(A)$ between the baryon states (see Eq.(12)). We will study the
cases of baryon octet and baryon decuplet respectively.

\vskip0.1in

\noindent {\bf 1) The baryon-octet case:} Noticing
SU(3)-multiplets decomposition formula
\begin{equation}
8\otimes8=27\oplus10^*\oplus8_F\oplus 8_D \oplus 1,
\end{equation}
we have (see Eq.(12)) $|n\rangle \in \{27,\; 10,\; 10^*,\;1\}$. AS
$|k\rangle \in$ nucleon, the nonzero ($n\neq k$) $\langle
n|D^{(8)}_{8,8}|k \rangle-$matrix elements are follows:
\begin{eqnarray}
\langle^{(10^*)}_{\mu\nu}|D^{(8)}_{(_{\;0}^{00})
(_{\;0}^{00})}|^{(8)}_{(_{\;{1\over 2}}^{{1\over 2}1})
(_{\;{1\over 2}}^{{1\over 2}1})}\rangle &=&\sqrt{1\over 20},
\\ \nonumber
\langle^{(27)}_{\mu\nu}|D^{(8)}_{(_{\;0}^{00})
(_{\;0}^{00})}|^{(8)}_{(_{\;{1\over 2}}^{{1\over 2}1})
(_{\;{1\over 2}}^{{1\over 2}1})}\rangle &=&\sqrt{3\over 50},
\end{eqnarray}
here equation (8) and  the SU(3) CG coefficients listed in
Ref.\cite{Swa} have been used. Since the spins of $|k\rangle $ and
$|n\rangle $ are same, from Eqs (5) and (10) we have
\begin{equation}
E^{(0)}_k-E^{(0)}_n={1\over 2}b^{-2}[C_2(k)-C_2(n)],
\end{equation}
where $C_2(k)$ denotes the Casimir operator for the
$k$-dimensional irreducible representation of SU(3). Noticing
$C_2(8)=3,\;\;C_2(10^*)=6$ and $C_2(28)=8$, then
\begin{eqnarray}
E^{(0)}_8-E^{(0)}_{10^*}&=&-{3\over 2}b^{-2},\\ \nonumber
E^{(0)}_8-E^{(0)}_{27}&=&-{5\over 2}b^{-2},
\end{eqnarray}
Combining the above equations with the known results in
Refs\cite{Gua} and \cite{Kan}, we have the nucleon's mass to the
second order in perturbation
\begin{equation}
M_N=M_8-{3\over 10}m-{43\over 750}g,
\end{equation}
where $M_s=\langle H_0\rangle_{\lambda=8}+m,\;\; g=m^2b^2$.
Through the similar computations, we can obtain the masses of
$\Lambda,\;\;\Sigma$ and $\Xi$ to the second order in perturbation
\begin{eqnarray}
M_\Lambda=M_8-{1\over 10}m-{9\over 250}g, \\ \nonumber
M_\Sigma=M_8+{1\over 10}m-{37\over 750}g, \\ \nonumber
M_\Xi=M_8+{1\over 5}m-{3\over 125}g,
\end{eqnarray}
where $M_8,\;m$ and $g$ are model-dependent variables. Eliminating
$M_8,\;m$ and $g$ in Eqs (17) and (18), we have
\begin{equation}
2(M_N+M_\Xi)=3M_\Lambda+M_\Sigma+{2\over
13}(M_N+M_\Sigma-2M_\Lambda).
\end{equation}
Comparing Eq.(19) with Eq.(1), we can see that the last term on
the right-hand side of Eq.(19) serves as a correction coming from
the skyrmion dynamics to GOR in the octet case. Thus we get the
desired result
\begin{equation}
\delta m_8={2\over 13}(M_N+M_\Sigma-2M_\Lambda).
\end{equation}
By using the mass relation of Eq.(19), the mass variable
$(M_N,\;M_\Sigma,\;M_\Lambda)$ in Eq.(20) can be changed to be
$(M_\Xi,\;M_\Sigma,\;M_\Lambda)$, or $(M_N,\;M_\Sigma,\;M_\Xi)$,
or $(M_N,\;M_\Xi,\;M_\Lambda)$. Namely, we can also write $\delta
m_8$ as
\begin{eqnarray}
\delta m_8&=&{1\over 12}(3M_\Sigma-2M_\Xi-M_\Lambda)\\ \nonumber
          &=&{2\over 35}(5M_\Sigma-M_N-M_\Xi)\\ \nonumber
          &=&{1\over 15}(6M_N+4M_\Xi-10M_\Lambda).
\end{eqnarray}
These four expressions of $\delta m_8$ are equivalent to each
other in principle. However, since equation (19) is still not an
exact identity of masses when experimental mass data are inputs,
the numerical results given by Eqs (20)--(21) we get
$$\delta m_8=-15.35\pm1.32 MeV.$$
 This second order correction to $\delta
m_8$ agrees with the experiment qualitatively, and in quantitative
respect it is compatible with the data showing in Eq (3) roughly.

\vskip0.1in

 \noindent {\bf 2) The baryon-decuplet case}: As $|k\rangle $ (in
 EQ. (12)) belongs to decuplet, from the SU(3)-decomposition
 formula $10\otimes 8=35\oplus 27 \oplus 10 \oplus 8$, we have $|n
 \rangle \in \{35,\;27,\;8\}$. Using again the formulas (8) and
 (15), the CG coefficients of SU(3)\cite{Swa} and $C_2(35)=12$, we
 get the masses of the baryon decuplet to the second order in
 perturbation
\begin{eqnarray}
M_\Delta=M_{10}-{1\over 8}m-{85\over 672}g,\hskip0.3in &&
M_{\Sigma^*}=M_{10}-{26\over 336}g, \\
M_{\Xi^*}=M_{10}+{1\over 8}m-{9\over 224}g,\hskip0.3in &&
M_{\Omega}=M_{10}+{1\over 4}m-{5\over 336}g,
\end{eqnarray}
where $M_{10}=\langle H_{10} \rangle_{\lambda=10}.$ Here there are
four equations and three unknowns. By solving these
over-determinant equations, a constrained equation is left, which
is
\begin{equation}
M_\Omega-M_\Delta=3(M_{\Xi^*}-M_{\Sigma^*}).
\end{equation}
Equation (24) is called the Okubo relationship which holds to the
second order in flavor breaking as shown by Okubo long
ago\cite{Oku} and by Morpurgo recently\cite{Mor}. Here, we reach
such conclusion again in the skyrmion formalism. Equations
(22)--(23) show that the equal spacing rule for the decuplet
(i.e., GOR in decuplet case) no longer holds. Using the
definitions of $\delta m^{(1)}_{10}$ and $\delta m^{(2)}_{10}$
(Eq.(2)), and the mass-splitting formulas (22)--(23), we have
$$\delta m^{(1)}_{10}={g\over 84},\hskip0.3in \delta m^{(2)}_{10}
={g\over 42}.$$ Then we get $\delta m^{(2)}_{10}/\delta
m^{(1)}_{10}=2$, which is in good agreement with the experiment
(see Eqs.(3)) $(\delta m^{(2)}_{10}/\delta m^{(1)}_{10})_{\rm
expt}=2.1$. From Eqs (22)--(23), $g$ can be found out to be $84(2
M_{\Sigma^*}-M_{\Xi^8}-M_\Delta)$. Then we get $\delta
m^{(1)}_{10}=2 M_{\Sigma^*}-M_{\Xi^8}-M_\Delta$. By using the
Okubo relationship (24) the mass variables in $\delta
m^{(1)}_{10}=2 M_{\Sigma^*}-M_{\Xi^8}-M_\Delta$ could be changed.
The other three expressions for $\delta m^{(1)}_{10}$ read
\begin{eqnarray}
\delta m_{10}^{(1)}&=&{1\over 3}(3M_{\Xi^*}-2M_\Omega-M_\Delta)\\
\nonumber
          &=&{1\over 3}(3M_{\Sigma^*}-2M_\Delta-M_\omega)\\ \nonumber
          &=&2M_{\Xi^*}-M_{\Sigma^*}-M_\Omega.
\end{eqnarray}
These four $\delta M_{10}^{(1)}$ expressions are equivalent in
principle, but their numerical results are not exactly same since
equation (24) is approximate just like the above octet case. Thus
we have
$$
\delta m_{10}^{(1)}=7.05\pm2.55 MeV, \hskip0.4in \delta
m_{10}^{(2)}=14.1\pm 5.1 MeV,
$$
where $\delta m^{(2)}_{10}/\delta m^{(1)}_{10}=2$ has been used.
They agree with the experimental data shown in Eqs (3).

The picture that the baryons are regarded as the chiral solitons
has extensively been investigated during last ten years\cite{Adk}.
It is well known that the quantum mechanics of SU(3) skyrmion is a
basic formalism to deal with $(ud)-$ and $s-$flavor breaking in
the model. Following the investigations of the first order in
flavor breaking in this quantum mechanics problem, we have
completed the calculations of the second order corrections to the
baryon-mass splitting and shown that the results support the
soliton pictures. Especially, since high order in perturbations of
quantum mechanics are intimately related to the $H'$-matrix
elements between various eigenstates of $H_0$ (see Eqs (9)--(12)),
our above results can be regarded as signals of the existence of
some SU(3) rotation excitation states of baryons: 27-let (with
spin ${1\over 2}$ or ${3\over 2}$, $10^*$-let (with spin ${1\over
2}$) and 35-let (with spin ${3\over 2}$. It is easy to be sure
that the wavefunctions of these SU(3) states satisfy the
constrained condition coming from the Wess-Zumino term in the QCD
effective Lagrangian, i.e., the spin hypercharge
$Y_R=1$\cite{Gua,Yab}. So they should be physical states in QCD.

From Eq.(5) the kinetic energy of usual SU(2) rotations of
skyrmion is
$T(SU(2))=[C_2(SU(2))]/(2I_2),\;I_2=(a^2b^2)/(a^2-b^2)$, where
$I_2$ is the moment of inertia for SU(2) rotation, and
$C_2(SU(2))\;(= \sum_{A=1}^3 R_AR_A\equiv \hat{J}^2)$ is the
Casimir operator for SU(2), or the square of total angular
momentum operators. Similarly, again from Eq.(5), the kinetic
energy of SU(3) rotations of skyrmion reads
$T(SU(3))=[C_2(SU(3))]/(2I_3),\;I_3=b^2$, where $I_3$ corresponds
to the monment of inertia for SU(3) rotation, $C_2(SU(3))\;(=
\sum_{i=1}^8 L_iL_i)$ is the Casimir operator for SU(3). Thus the
mode of SU(3) rotation excitations can be understood as a natural
extension of the usual SU(2) rotation excitations. Clearly, all of
the quantum numbers for both SU(3) rotation excitation state
(SU(3)RES) and its corresponding octet or decuplet baryons are the
same except their SU(3) Casimir operators. By using $10^*_x,\;27_x
$ and $35_x$ to denote SU(3)RES of $x\in \{{\rm
octet,\;decuplet}\}$, we may reasonably conclude that there exist
16 SU(3)RES which have visible contributions to the corrections to
GOR, and they are
$10^*_N,\;10^*_\Lambda,\;10^*_\Sigma,\;10^*_\Xi,\;27_N,\;27_\Lambda,\;
27_\Sigma,\;27_\Xi,\;27_\Delta,\;27_{\Sigma^*},\;
27_{\Xi^*},\;27_\Omega,\;35_\Delta,\;35_{\Sigma^*},\; 35_{\Xi^*}$
and $ 35_\Omega$.

Finally, we would like to make two remarks. First, in the present
paper the mass spectra of the octet case and the decuplet case
calculated in perturbations respectively and independently.
However, due to the limitations of perturbation method for the
$(ud)-$ and $s-$flavor symmetry breaking, it cannot be extended to
the case of the octet and the decuplet combined. If one wants to
get a complete mass spectrum (including octet and decuplet
simultaneously), the skyrmion quantum mechanics problem in the
SU(3) collective coordinate space (see Eqs (4)--(6)) have to be
solved exactly. Unfortunately, this goal can merely be reached
numerically\cite{Yab}, instead of analytically. Therefore, in
order to reveal the physical implications explicitly, our
analytical analyses in the above discussions are necessary.
Second, our analyses on $\delta m_8,\;\delta m_{10}^{(1)}$, and
$\delta m_{10}^{(2)}$ are model-independent. In other words, we
did not use any concrete ${\cal L}_{eff}$ to do calculations till
now, and all conclusions in this paper come from the studies on
generic structure of the Hamiltonian of the SU(3) quantum
mechanics (Eqs (4)--(6)). Thus our predictions on existence of
SU(3)RES are also model-independent in the skyrmion formalism.

\end{document}